\documentclass[preprint,showkeys,unsortedaddress,superscriptaddress, amsfonts,showpacs,amsmath,amssymb]{revtex4}
\usepackage[dvips]{color}
\usepackage{epsfig}
\usepackage{amsmath}
\usepackage{graphicx}
\def\Box{\hbox{$\rlap{$\sqcup$}\sqcap$}}

\begin{document}

\title{\bf Chameleonic Generalized Brans--Dicke model and late-time acceleration }

\author{Hossein Farajollahi}
\email{hosseinf@guilan.ac.ir}
\affiliation{Department of Physics, University of Guilan, Rasht, Iran}

\author{Mehrdad Farhoudi}
\email{m-farhoudi@sbu.ac.ir}
\affiliation{Department of Physics, Shahid Beheshti University, G.C., Evin, Tehran 1983963113, Iran}

\author{Amin Salehi}
\email{a.salehi@guilan.ac.ir}
\affiliation{Department of Physics, University of Guilan, Rasht, Iran}

\author{Hossein Shojaie}
\email{h-shojaie@sbu.ac.ir}
\affiliation{Department of Physics, Shahid Beheshti University, G.C., Evin, Tehran 1983963113, Iran}
\date{\today}

\begin{abstract}

In this paper we consider Chameleonic Generalized Brans--Dicke Cosmology in the framework of FRW universes. The bouncing solution and phantom crossing is investigated for the model. Two independent cosmological tests: Cosmological Redshift Drift (CRD) and distance modulus are applied to test the model with the observation.

\end{abstract}



\maketitle

\section{Introduction}

Various cosmological observations, mainly Cosmic Microwave Background (CMB)~\cite{WMAP,Komatsu}, Supernova type Ia (SNIa)~\cite{SNIa,Riess}, Weak Lensing~\cite{WeakLensing}, Baryon Acoustic Oscillations (BAO)~\cite{BAO}, and redshift surveys such as 2dF Galaxy Redshift Survey (2dFGRS)~\cite{2dFGRS} at low redshift and DEEP2 redshift survey~\cite{DEEP2} at high redshift, have provided cross-checked data to determine cosmological parameters with high precision. These parameters imply that our approximately $13.72$ Gyear-old universe is nearly spatially flat, homogeneous and isotropic at large scale, i.e. a Friedmann-Lema\^{\i}tre-Robertson-Walker (FLRW) universe with zero curvature, and has entered an accelerating phase since $z\approx0.46$~\cite{Riess}. Moreover, according to this so-called \emph{Concordance Model} (or $\Lambda$CDM model), the universe consists of $4.6\%$ baryonic matter, $22.8\%$ non-relativistic unknown matter, namely dark matter (DM), and a remarkable amount of $72.6\%$ smoothly distributed dominant component, dubbed dark energy (DE)~\cite{Komatsu}.

The equation of state (EoS), $w=p/\rho$, of DE, is the main parameter which determines the gravitational effect of DE on the evolution of the universe, and can be measured from observations without need to have a definite model of DE. Strong evidences imply that the EoS of DE lies in a narrow range around $w\approx-1$ and has a smooth evolution~\cite{Riess,Amanullah}. However, slow variation of $w$ in cosmic time is not excluded. Theoretically, one can classify the EoS of DE with respect to the barrier $w=-1$~\cite{Cai}, namely the phantom divide line (PDL). That is, DE with the EoS of $w=-1$ employs the cosmological constant, $\Lambda$, with a constant energy density. The case with dynamical EoS where $w\geq-1$, is referred to as quintessence~\cite{quintessence} and $w\leq-1$ corresponds to an odd theoretical matter case, known as phantom energy~\cite{phantom,Caldwell}, which potentially has elusive properties. For instance, phantom energy density increases with time as the universe expands, which consequently leads to divergence of the scale factor in finite time~\cite{Caldwell}. Moreover, sound speed in this kind of matter \emph{may} be superluminal. In addition to these three subclasses, there is another suggestion for the EoS of DE, the so-called quintom~\cite{quintom}, in which $w$ crosses the phantom divide line as time evolves. It behaves like quintessence in some periods of history of the universe, and like phantom in the rest. It is worth noting that a single scalar field with conventional features minimally coupled to gravity cannot cross the barrier according to a corresponding no-go theorem~\cite{no-go}. For a complete discussion about the origin of DE, see for example Refs.~\cite{Cai,DEreview} and references therein.

In addition to the above classification according to EoS, one can consider different models for DE. Actually, there is a wide variety of models concerning dynamics for DE; for instance, quintessence models which exploit canonical fields with dynamical EoS~\cite{quintessence}, k-essence models with a non-canonical kinetic term~\cite{k-essence}, quintom models usually with two scalar fields, one as a quintessence and one as a phantom~\cite{quintom}, Chaplygin gas models which unify dark matter and dark energy via a superfluid~\cite{Chaplygin} , tachyon fields motivated by string theory~\cite{tachyon}, chameleonic fields in which a scalar field is coupled to matter~\cite{chameleon,Khoury-Brax}, and modified gravity models regarded as modifications to Einstein-Hilbert action~\cite{modified-gravity}.

The modified gravity models are obtained in different ways, for example, by replacing Ricci scalar, $R$, by $f(R)$  as the well-known $f(R)$ modified gravity~\cite{f(R)}, by adding additional scalar terms like the Gauss-Bonnet term~\cite{Gauss-Bonnet}, or by non-minimally coupling of a scalar field to the Ricci scalar such as the Brans-Dicke (BD) theory of gravitation~\cite{BD} or its generalization (GBD) by employing the BD parameter, as a function of cosmic time.

To explain the late time acceleration of the
universe as mentioned it is most often the case that the scalar fields interact with matter; directly due to a matter Lagrangian
coupling, indirectly through a coupling to the Ricci scalar or as the result of quantum loop corrections \cite{Damouri}--\cite{Biswass}. If the
scalar field self-interactions are negligible, then the experimental bounds on such a field are very strong; requiring it to either
couple to matter much more weakly than gravity does, or to be very heavy \cite{Uzan}--\cite{Damourm}. Unfortunately, such scalar field is usually very light and its coupling to matter should be tuned to
extremely to small values in order not to be conflict with the Equivalence Principal \cite{nojiri}.

An alternative attempt to overcome the problem with light scalar fields has
been suggested in chameleon cosmology \cite{Khoury-Brax}\cite{Mota}. In
the proposed  model, a scalar field couples to matter with gravitational strength, in harmony with general expectations from
string theory whilst at the same time remaining very light on cosmological scales. The scalar field which is very light on cosmological scales is permitted
to couple to matter much more strongly than gravity does, and yet still satisfies the current experimental and observational
constraints. The cosmological value of such a field evolves over Hubble time-scales and could potentially cause the late-time acceleration of our Universe \cite{Brax2}. The crucial feature that these models possess are that the mass of the scalar field depends on the
local background matter density. While the idea of a density-dependent mass term is not new \cite{Wett}--\cite{Mot}, the
work presented in \cite{Khoury-Brax} \cite{Brax2} is novel in that the scalar field can couple directly to matter with gravitational strength.

In this work in addition to the most expectations presented in~\cite{Khoury-Brax}, we investigate an integration of GBD and chameleon models, calling it as chameleonic generalized Brans-Dicke model (CGBD), which allows the scalar field to couple with both gravity and matter and the BD parameter to be a function of the scalar field. Since the scalar field, which treats very lightly on large scales, can couple to matter much more stronger than gravity does, the observational constraints can be satisfied by this model. As mentioned, to construct a model which its EoS can smoothly cross the PDL, it is necessary to add unconditional features to scalar field in order to have extra degrees of freedom. We show that the CGBD model studied in this work is capable to predict not only a PDL crossing scenario but also a bouncing behavior for the universe.

To directly measure the expansion history of the universe, it is necessary to find a probe which is model-independent and is based on fewer assumptions. This can be performed by focusing on dynamics of homogeneous and isotropic large scale features of the universe, that is, the Robertson-Walker (RW) geometry. An example of such a probe is the cosmological redshift drift (CRD) first proposed by Sandage and McVittie~\cite{Sandage-McVittie}. This method~\cite{CRD} measures the drift of redshift of specific absorption lines, specially  those arising from the Lyman-$\alpha$ transition in the spectra of high-redshift galaxies and quasars (QSOs) in a time interval of several years~\cite{Liske}. The drift, $\dot z$, maps the Hubble parameter, $H(z)$, and consequently determines acceleration or deceleration of the universe, and only vanishes when the expansion rate of the scale factor is constant. However, for varying expansion, it is of order of the Hubble parameter, approximately $10^{-10}$, and gives rise to an acceleration of order of few centimeters per second per year, and can be swamped by peculiar acceleration of cosmological objects. A second probe is the observations of the difference between the distance modulus –- red-shift relation for SNIa that verifies the late-time accelerated expansion of the universe and will be discussed in here.

The manuscript is arranged as follows. In the next section, we introduce the CGBD model and analytically study conditions for the EoS parameter to pass the PDL. In Section~III, by numerical calculation a possible bouncing of the universe and EoS parameter crossing is examined. In Section~IV, the inclusion of observational data in our model through CRD and the difference in the Distance Modulus ($\mu(z)$) tests are investigated. Section~V represents the results and summary.

\section{The CGBD model}

We consider the CGBD gravity in the presence of cold dark matter with the action given by,
\begin{eqnarray}\label{ac}
S=\int d^{4}x\sqrt{-g}\left(\phi
R-\frac{\omega(\phi)}{\phi}\phi_{,\mu}\phi^{,\mu}+f(\phi)L_{m}\right),
\end{eqnarray}
where $R$ is the Ricci scalar and
$\omega(\phi)$ is the scalar field dependent BD parameter.
The matter Lagrangian is modified as $f(\phi)L_{m}$, where $f(\phi)$ is an
analytic function of the scalar field. This term represents the nonminimal interaction between the matter and the scalar field.
Metric variation of action (\ref{ac}) gives,
\begin{eqnarray}\label{fieldeq}
G_{\mu\nu}=\frac{\omega(\phi)}{\phi^{2}}\left(\phi_{,\mu}\phi_{,\nu}
-\frac{1}{2}g_{\mu\nu}\phi_{,\rho}\phi^{,\rho}\right)
+\frac{1}{\phi}\left(\phi_{,\mu;\nu}-g_{\mu\nu}\Box\phi\right)+\frac{f(\phi)}{\phi}T_{\mu\nu}.
\end{eqnarray}
In a flat FLRW cosmology, field equations (\ref{fieldeq}) become
\begin{eqnarray}\label{fried1}
3\frac{\dot{a}^{2}}{a^{2}}=\frac{\rho_{m}f(\phi)}{\phi}-3\frac{\dot{a}}{a}\frac{\dot{\phi}}{\phi}
+\frac{\omega(\phi)}{2}\frac{\dot{\phi}^{2}}{\phi^{2}},
\end{eqnarray}
and
\begin{eqnarray}\label{fried2}
2\frac{\ddot{a}}{a}+\frac{\dot{a}^{2}}{a^{2}}=-\frac{p_{m}f(\phi)}{\phi}-2\frac{\dot{a}}{a}\frac{\dot{\phi}}{\phi}
-\frac{\omega(\phi)}{2}\frac{\dot{\phi}^{2}}{\phi^{2}}-\frac{\ddot{\phi}}{\phi}.
\end{eqnarray}
 On the other hand, the variation of action (\ref{ac}) with respect to the spatially homogeneous scalar field $\phi(t)$ yields,
\begin{eqnarray}\label{phiequation}
\ddot{\phi}+3\frac{\dot{a}}{a}\dot{\phi}=\frac{1}{3+2\omega(\phi)}\left[(\rho_m-3p_m)\left(f(\phi)-\frac{1}{2}\phi f'(\phi)\right)-\omega'(\phi)\frac{\dot{\phi}^{2}}{\phi}\right],
\end{eqnarray}
where the prime denotes derivative with respect to $\phi$. In the above equations, $\rho_m$ and $p_m$ are respectively energy density and isotropic pressure of the barotropic perfect fluid. The EoS of the baryonic and dark matter is $p_{m}=w_m\rho_{m}$. From equations (\ref{fried1})--(\ref{phiequation}) , one can easily derive the modified conservation equation
\begin{eqnarray}\label{conserv}
\frac{\dot\rho_m}{\rho_m}+3H(1+w_m)=-3w_m\frac{\dot{f}}{f},
\end{eqnarray}
which readily integrates to yield
\begin{eqnarray}\label{density}
\rho_{m}=\frac{M}{a^{3(1+w_{m})}f^{3w_{m}}}
\end{eqnarray}
where $M$ is a constant of integration. The equation $(\ref{density})$ shows that for $f=1$ ,$\rho_{m}=\frac{M}{a^{3(1+w_{m})}}$ as expected $\rho_{m}$ decreases with increasing $f(\phi)$.
In comparison with the standard Friedmann
equations one may identify $\rho_{\rm_{eff}}$ and $p_{\rm_{eff}}$ from equations (\ref{fried1})
and (\ref{fried2}) as
\begin{eqnarray}\label{rhoeff}
3H^2=\rho_{\rm_{eff}}\equiv
\frac{\rho_mf}{\phi}-\frac{\dot{\phi}}{\phi}(3H-\frac{\omega}{2}\frac{\dot\phi}{\phi}),
\end{eqnarray}
and
\begin{eqnarray}\label{peff}
-\left(2\dot{H}+3H^2\right)=p_{\rm_{eff}}&\equiv
&\frac{p_mf}{\phi}+\frac{1-3w_m}{3+2\omega}\frac{\rho_m}{\phi}(f-\frac{1}{2}\phi f')\\&+&\frac{\dot{\phi}}{\phi}
(-H+\frac{\omega}{2}\frac{\dot\phi}{\phi}-\frac{\omega'}{3+2\omega}\frac{\dot\phi}{\phi}).\nonumber
\end{eqnarray}
where we have supposed $p_{\rm_{eff}}=w\rho_{\rm_{eff}}$. Consequently conservation equation reads as
\begin{eqnarray}\label{cons}
\dot\rho_{\rm_{eff}}+3H(1+w)\rho_{\rm_{eff}}=0.
\end{eqnarray}
From equation (\ref{fried1}) we obtain,
\begin{eqnarray}\label{h00}
H=-\frac{\dot\phi}{2\phi}\pm\sqrt{\frac{\rho_mf}{3\phi}+\frac{3+2\omega}{12}\frac{\dot\phi^2}{\phi^2}}.
\end{eqnarray}
One can rewrite the second Friedmann equation, (\ref{fried2}), as
\begin{eqnarray}\label{Hdot}
\dot{H}=\frac{1}{2}[H\frac{\dot{\phi}}{\phi}-\omega\frac{\dot{\phi^{2}}}{\phi^{2}}
-\frac{\ddot{\phi}}{\phi}-\frac{(1+3w_m)\rho_{m}f}{\phi}].
\end{eqnarray}
Differentiating~(\ref{Hdot}) with respect to time gives,
\begin{eqnarray}\label{Hddot}
\ddot{H}=\frac{\dot{\phi}}{2\phi}\Big[\frac{3\omega}{2}\frac{\dot{\phi}^{2}}{\phi^{2}}-
(\dot{\omega}+\frac{H}{2})\frac{\dot{\phi}}{\phi}-(2\omega-\frac{1}{2})\frac{\ddot{\phi}}{\phi}+\frac{(1+3w_m)}{2}\rho_{m}f\Big]
-\frac{\dddot{\phi}}{2\phi}-\frac{(1+3w_m)\dot{(\rho_{m}f)}}{2\phi}.
\end{eqnarray}
To have the PDL crossing, that is $\dot{w}\neq0$ as $w\rightarrow-1$, one should have $\dot{H}=-\frac{1}{2}(1+w)\rho_{\rm_{eff}}=0$ and $\ddot H=-\frac{1}{2}\dot w\rho_{\rm_{eff}}-\frac{1}{2}(1+w)\dot\rho_{\rm_{eff}}\neq0$ when $w=-1$. From equation ~(\ref{Hddot}), we observe that it is always possible to have non vanishing $\ddot{H}$ and also vanishing $\dot{H}$ when $w=-1$, if one of the following conditions is satisfied:

1) $\dot{\phi}\neq 0$, $\dddot{\phi} = 0$, $w_m = -1/3$, $\dot{(\rho_{m}f)} =  0$,

2) $\dddot{\phi}\neq 0 $, $\dot{\phi}= 0$, $w_m = -1/3$, $\dot{(\rho_{m}f)} =  0$,

3) $w_m\neq -1/3$, $\dot{\phi}= 0$, $\dddot{\phi} = 0$, $\dot{(\rho_{m}f)} =  0$,

4) $\dot{(\rho_{m}f)}\neq 0$, $\dot{\phi}= 0$, $\dddot{\phi} = 0$, $w_m = -1/3$.

In the first case, (1), $H(t)$ is given by equation (\ref{h00}), and
\begin{eqnarray}\label{Hdotb}
\dot{H}=\frac{1}{2}[H\frac{\dot{\phi}}{\phi}-\omega\frac{\dot{\phi^{2}}}{\phi^{2}}-\frac{\ddot{\phi}}{\phi}]=0.
\end{eqnarray}
By imposing the condition $\dot{H}$, we obtain,
\begin{eqnarray}\label{hddota}
\ddot{H}&=&\frac{\dot{\phi}^2}{2\phi^2}\Big[\omega(1+2\omega)\frac{\dot{\phi}}{\phi}-
(\dot{\omega}+2\omega H)\Big]\cdot
\end{eqnarray}
Therefore the conditions for having PDL crossing is that i) $\omega \neq 0$, ii) $\omega \neq -1/2$, or iii) $\dot{\omega} \neq 2\omega H $ and also the expression inside the bracket does not vanish. In addition, we need $\phi \neq 0$.

In the second case, (2), $H(t)=\pm\sqrt{\frac{\rho_mf}{3\phi}}$, $\dot{H}=\frac{-\ddot{\phi}}{2\phi}$ and $\ddot{H}=-\frac{\dddot{\phi}}{2\phi}$. The conditions for having PDL crossing is that $\ddot{\phi}=0$ and $\dddot{\phi}\neq 0$ when $w\rightarrow-1$. In addition, we need $\phi \neq 0$. In case (3), $H(t)$ is the same as case (2), $\dot{H}=\frac{-1}{2}[\frac{\ddot{\phi}}{\phi}+\frac{(1+3w_m)\rho_{m}f}{\phi}]$ and $\ddot{H}=0$. In this case since one of the PDL crossing conditions is not satisfied, the crossing does not occur. In case (4) again since $\ddot{H}=0$, the crossing does not occur.

Finally, the Hubble parameter as a function of redshift, $H(z)$, is obtained using~(\ref{rhoeff}) and~(\ref{cons}) as
\begin{eqnarray}\label{Hz}
H^2(z)=e^{3\int_0^z\frac{1+w(z')}{1+z'}dz'},
\end{eqnarray}
which by numerical methods can be used to trace the Hubble parameter versus redshift. Evidently, combination of the result of this model-dependent relation with the model-free observational data such as CRD and distance modulus, can be used for verification of model. With the analytic discussion for the PDL crossing given in this section, a numerical computation will be performed in the following section to illustrate the bouncing and PDL crossing dynamics for the model.

\section{Bouncing Behavior and PDL Crossing}

We explore the possible bouncing of the universe. To examine this, choose $t=0$ to be the bouncing point. Our effective energy density and pressure, equations (\ref{rhoeff}), (\ref{peff}) and consequently derived $\omega$ shows that it approaches negative values less than $-1$ in two periods of time.
Also by numerically solving the set of friedmann equations and scalar field equation we obtain $H(t)$ that provides a dynamical universe with
contraction for $t<0$, bouncing at $t=0$ and then expansion for $t>0$. The above
analysis clearly can be seen in the numerical calculation given in Fig.~1.

\begin{tabular*}{2.5 cm}{cc}
\includegraphics[scale=.4]{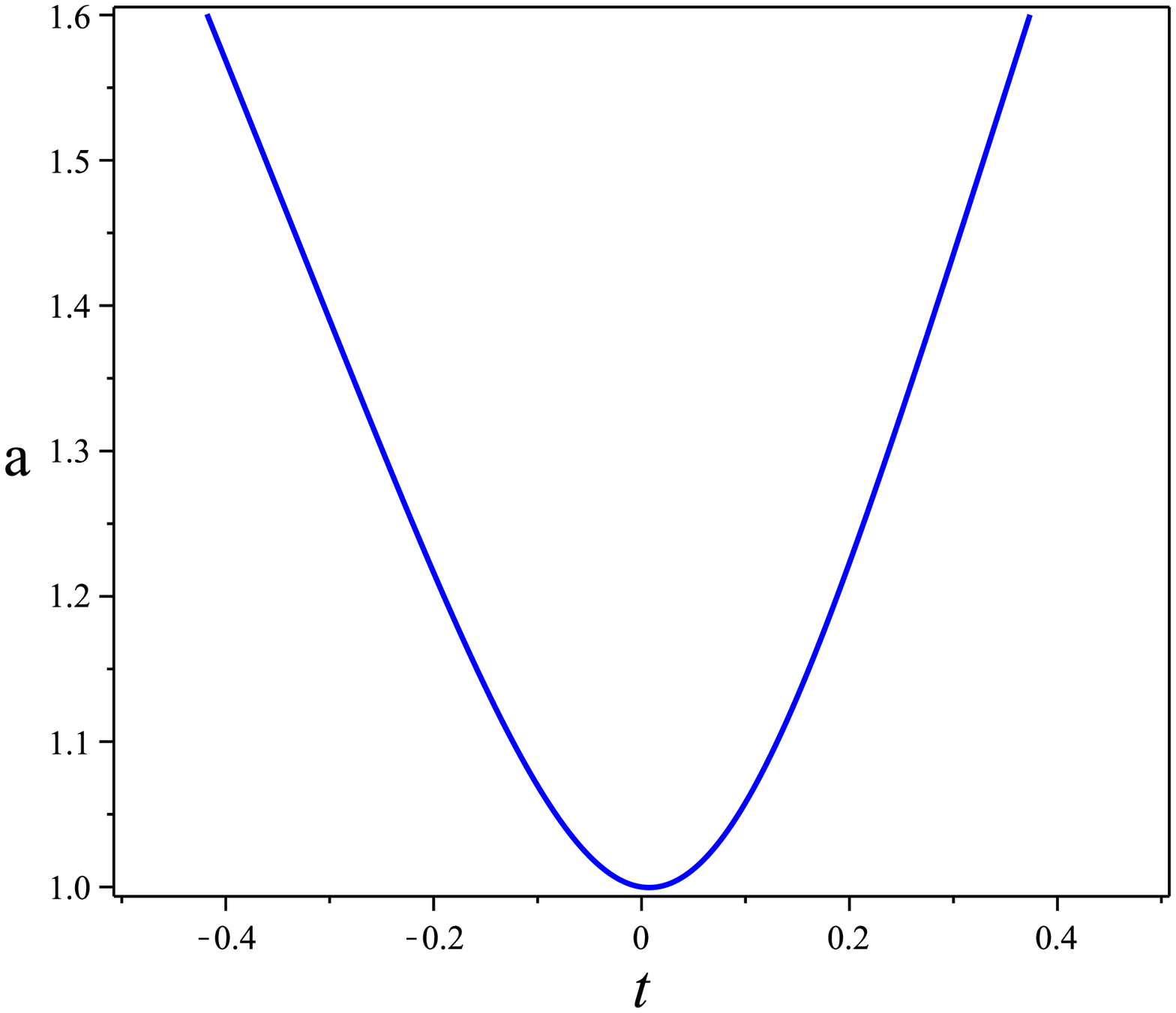}\hspace{0.1 cm}\includegraphics[scale=.4]{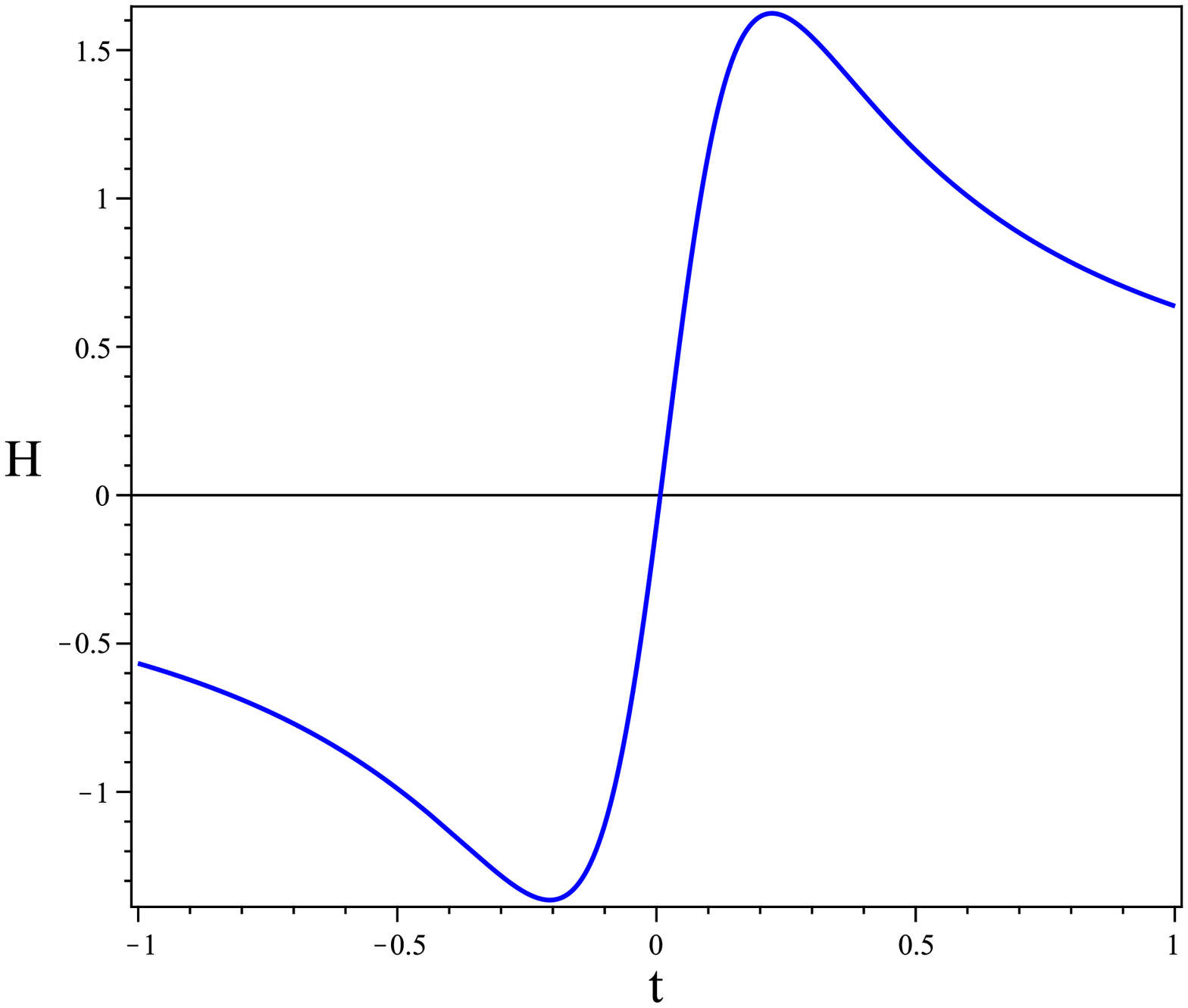}\hspace{0.1 cm}\\
Fig. 1:  The graphs of the scale factor $a$ and the Hubble parameter $H$, plotted as functions\\
of time for $\omega=\omega_{0}\phi(t)^{n}$, $\omega_{0} = -1000$, $n=-2$, $f=f_{0}e^{b\phi(t)}$,
$f_{0} = -2000$, and $b = -1$. \\
Initial values are $\phi(0)=-2$ , $\dot{\phi}(0)=1$ and $\dot{a}(0)=0.1$.\\
\end{tabular*}\\

\begin{tabular*}{2.5 cm}{cc}
\includegraphics[scale=.4]{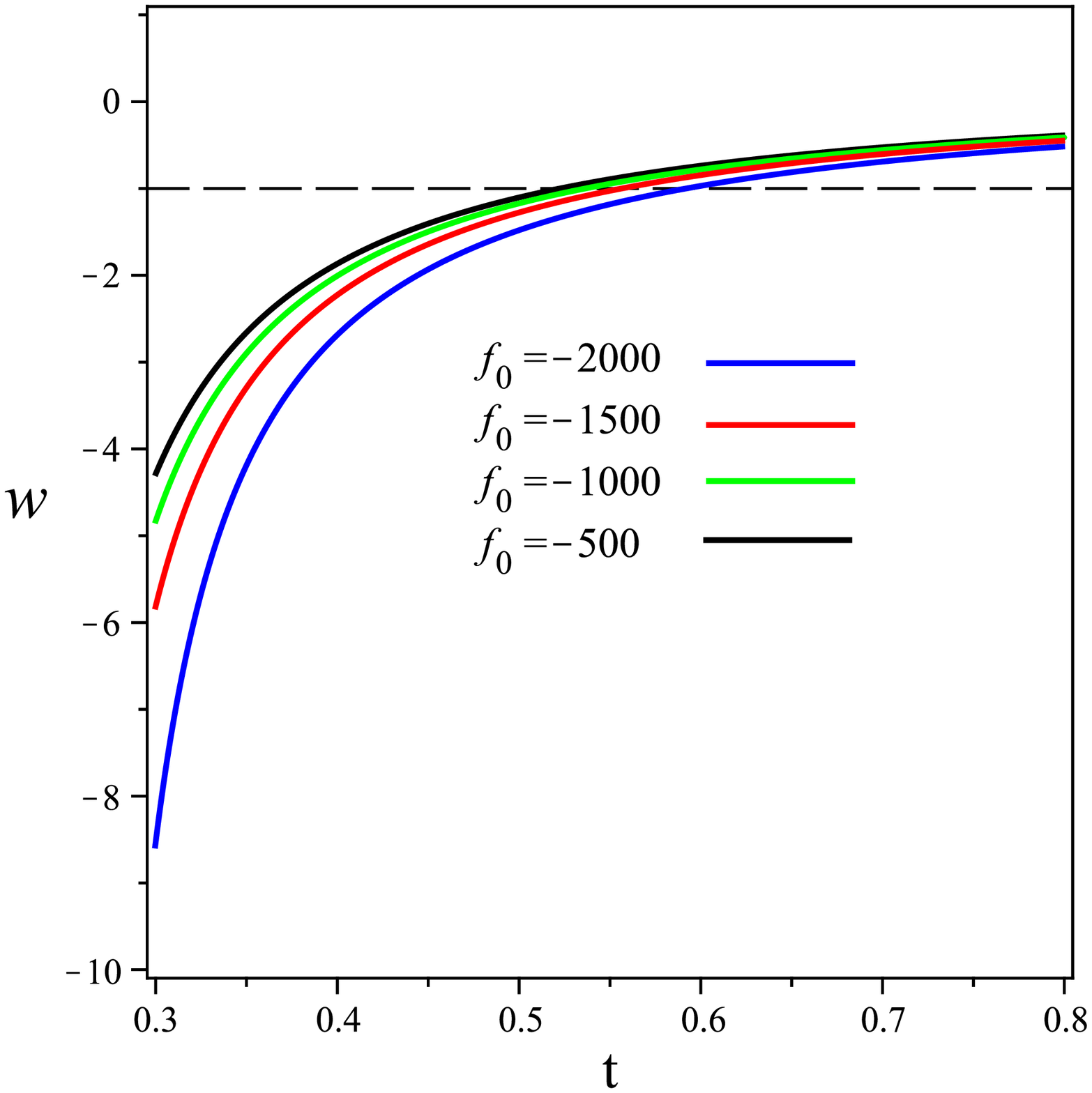}\hspace{0.1 cm}\includegraphics[scale=.4]{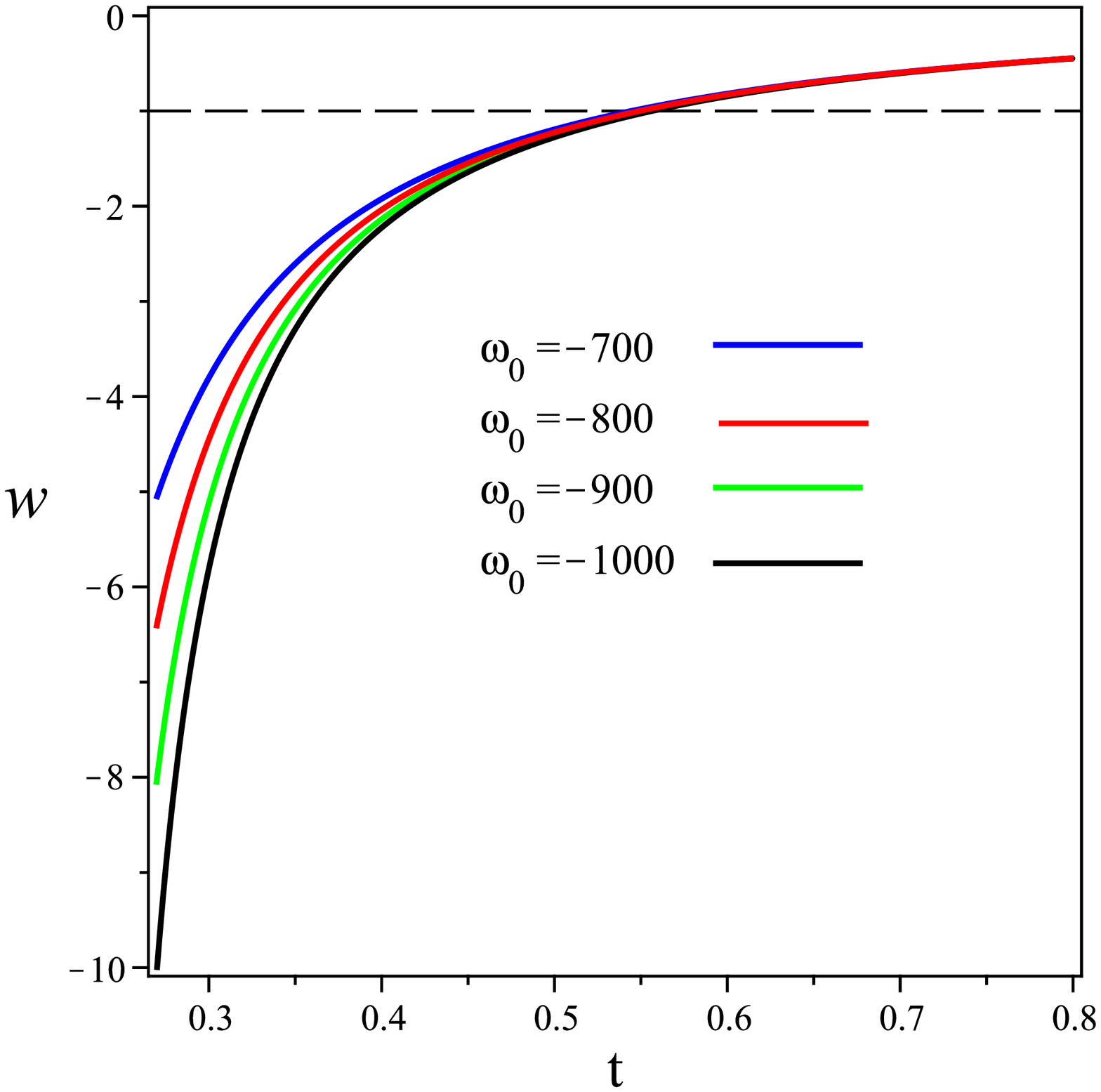}\hspace{0.1 cm}\\
Fig. 2:  The graphs of EoS parameter $w$, plotted as functions of time for\\
$\omega=\omega_{0}\phi(t)^{n}$, $n=-2$, $f=f_{0}e^{b\phi(t)}$, $b = -1$.
For $\omega_{0}=-1000$ and four different\\ values of $f_{0}$ (left panel) and for $f_{0}=-1000$ and four different values of $\omega_{0}$ (right panel).\\ Initial values are $\phi(0)=-2$, $\dot{\phi}(0)=1$, $\dot{a}(0)=0.1$.\\
\end{tabular*}\\

As Fig.~2 shows, in our model, the EoS parameter crosses PDL with $\dot w>0$, at different times, depending on the initial values and model parametrization. In addition, with respect to the bouncing,  during the contracting phase, the
scale factor $a(t)$ is decreasing, that is, $\dot{a}<0$, and in the
expanding phase we have $\dot{a}>0$. At the bouncing point,
$\dot{a}=0$, and so around this point $\ddot{a}>0 $ for a
period of time. Equivalently, in the bouncing cosmology, the Hubble
parameter $H$ runs across zero from $H<0$ to $H>0$, and at bouncing point $H=0$. This shows that a successful bounce requires $\dot H>0$ around the bouncing point. According to equation~(\ref{Hdot}), this result is equivalent to
\begin{eqnarray}\label{hdot1}
\dot{H}=-\frac{1}{2}(1+w)\rho_{\rm_{eff}}>0.
\end{eqnarray}
From Fig.~2, we see that at $t\rightarrow 0$, $w<-1$ and
$\dot{H}>0$ which satisfy the above condition. One of the well known problems in standard cosmology is the big bang singularity where the universe with zero volume has infinite density. We see that in our model at the bouncing point the scale factor $a(t)$ is
not zero and thus we avoid singularity faced in the standard cosmology.

 We note that Fig. 2 just shows the dynamics of the EoS parameter as a function of time.
At early time, crossing from the phantom phase to the non phantom phase towards $w=0$ looks like a transition say from inflation era to matter dominated era in early universe. The model does not show otherwise. For a more observationally oriented perspective, in the next section two cosmological tests for observable quantities are performed in terms of the redshift.

\section{The Cosmological Tests}

\subsection{CRD Test}

As mentioned in section~II, the relation~(\ref{Hz}) maps the expansion history of the universe for a given EoS. On the other hand, the effective EoS lies in a narrow range around $w=-1$. Following~\cite{CPL}, in Chevallier-Polarski-Linder (CPL) parametrization model one can use linear approximation,
\begin{eqnarray}\label{hdot1}
w_{cpl}\approx w_0-\frac{dw_{cpl}}{da}(a-1)=w_0+w_1\frac{z}{1+z},
\end{eqnarray}
where $w_0$ is current value of the effective EoS, and $w_1=-\frac{dw_{cpl}}{da}$ is its running factor. Using Eq. (\ref{hdot1}) we can obtain the following equation for Hubble parameter
\begin{equation}
\frac{H(z)^{2}}{H^{2}_{0}} =\Omega_m(1+z)^3+(1-\Omega_m)(1+z)^{3(1+w_0+w_1)}\times  \exp{\left[-3w_1(\frac{z}{1+z})\right]}\cdot \label{Hr}
 \end{equation}
In CPL model the parametrization is fitted for different values of $w_0$, $w_1$ and $\Omega_m$. In addition, in FLRW geometry, the CRD can be extracted from
\begin{eqnarray}\label{dotz}
\dot{z}=(1+z)H_0-H(z),
\end{eqnarray}
where immediately leads to velocity drift
\begin{eqnarray}\label{vdrift}
\dot{v}=cH_0-\frac{cH(z)}{1+z}.
\end{eqnarray}
The velocity drift with respect to the source for CPL model against
redshift is shown in Fig. 3 \cite{data}. A comparison of the model with the observational data shows that the best fit values are for $w_0=-2$ and $w_1=3.5$ in the region $z>3$.\\

 \begin{tabular*}{2.5 cm}{cc}
\includegraphics[scale=.4]{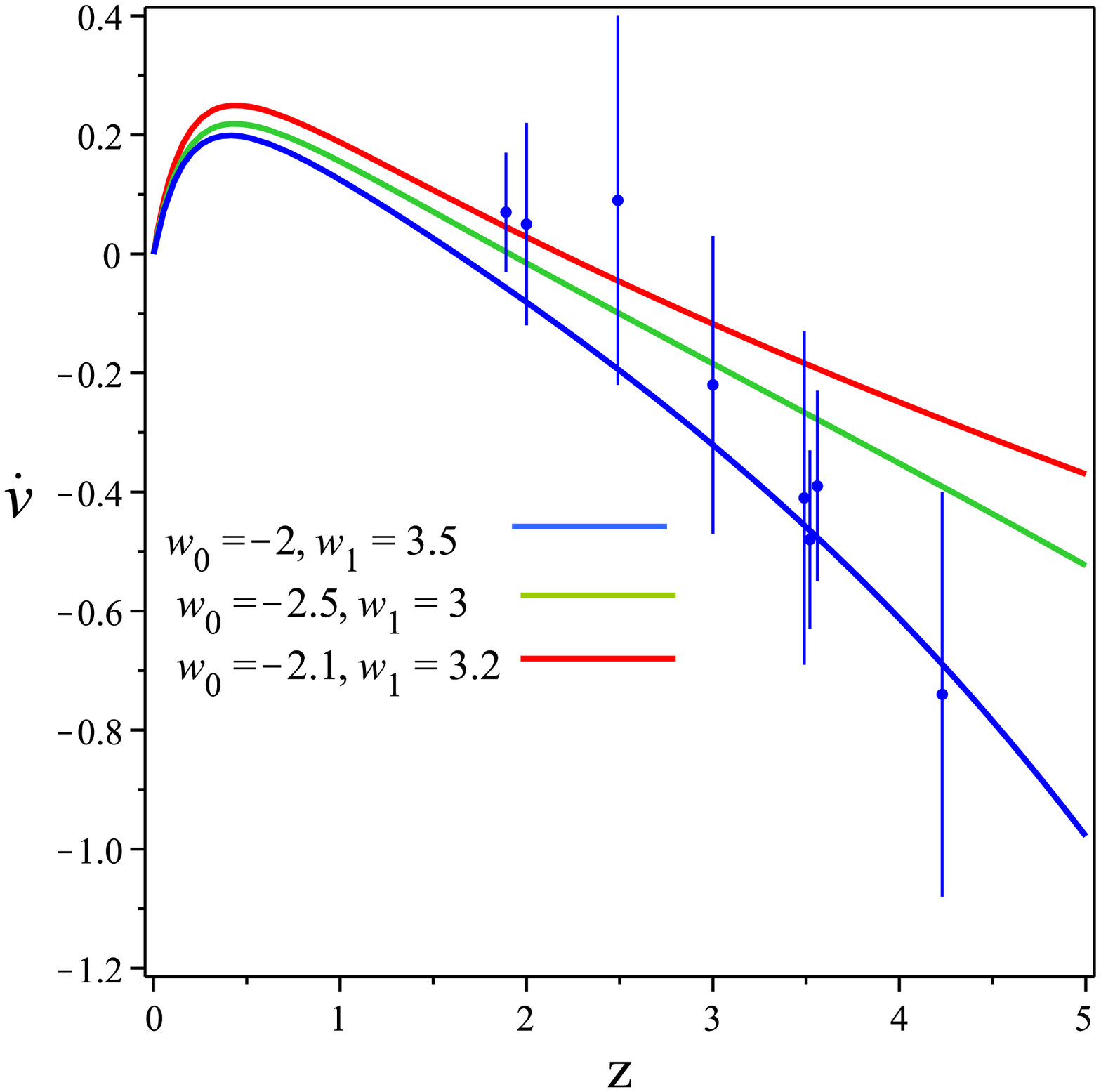}\hspace{0.1 cm}\\
Fig. 3: The graph of the CPL parameterization model in comparison with\\ the observational data, plotted as functions
of $z$ for different values of $w_0$ and $w_1$\\
\end{tabular*}\\

To draw the left panels of Figs.~4,~5 and ~6, we have used the effective EoS parameter $w$ for some plausible functions $f(\phi)$ and $\omega(\phi)$. The right panel of Fig.~4 shows the velocity drift against $z$
for different values of $f_0$, with respect to the observational data~\cite{data,Liske}.

\begin{tabular*}{2.5 cm}{cc}
\includegraphics[scale=.4]{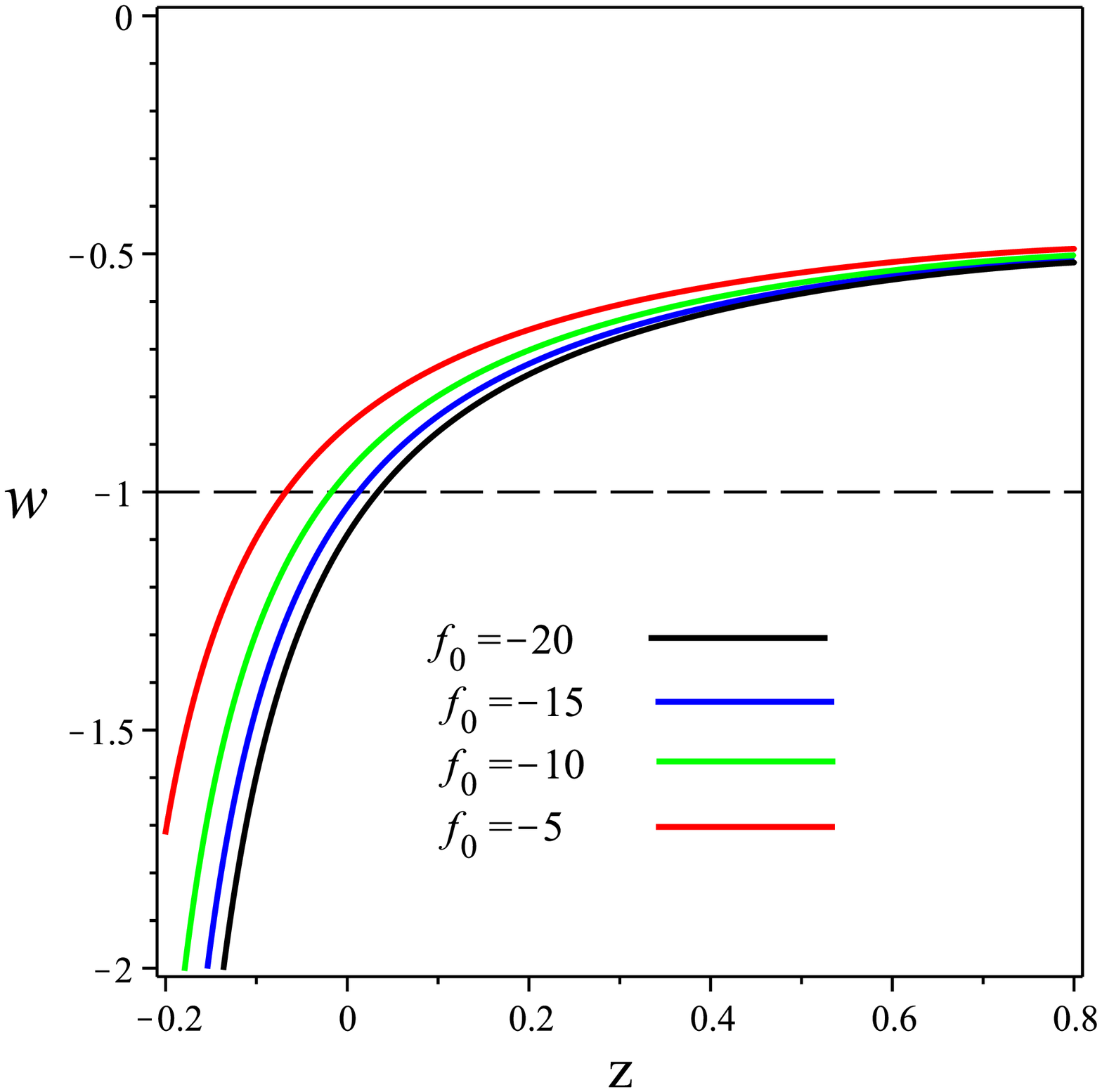}\hspace{0.1 cm}\includegraphics[scale=.4]{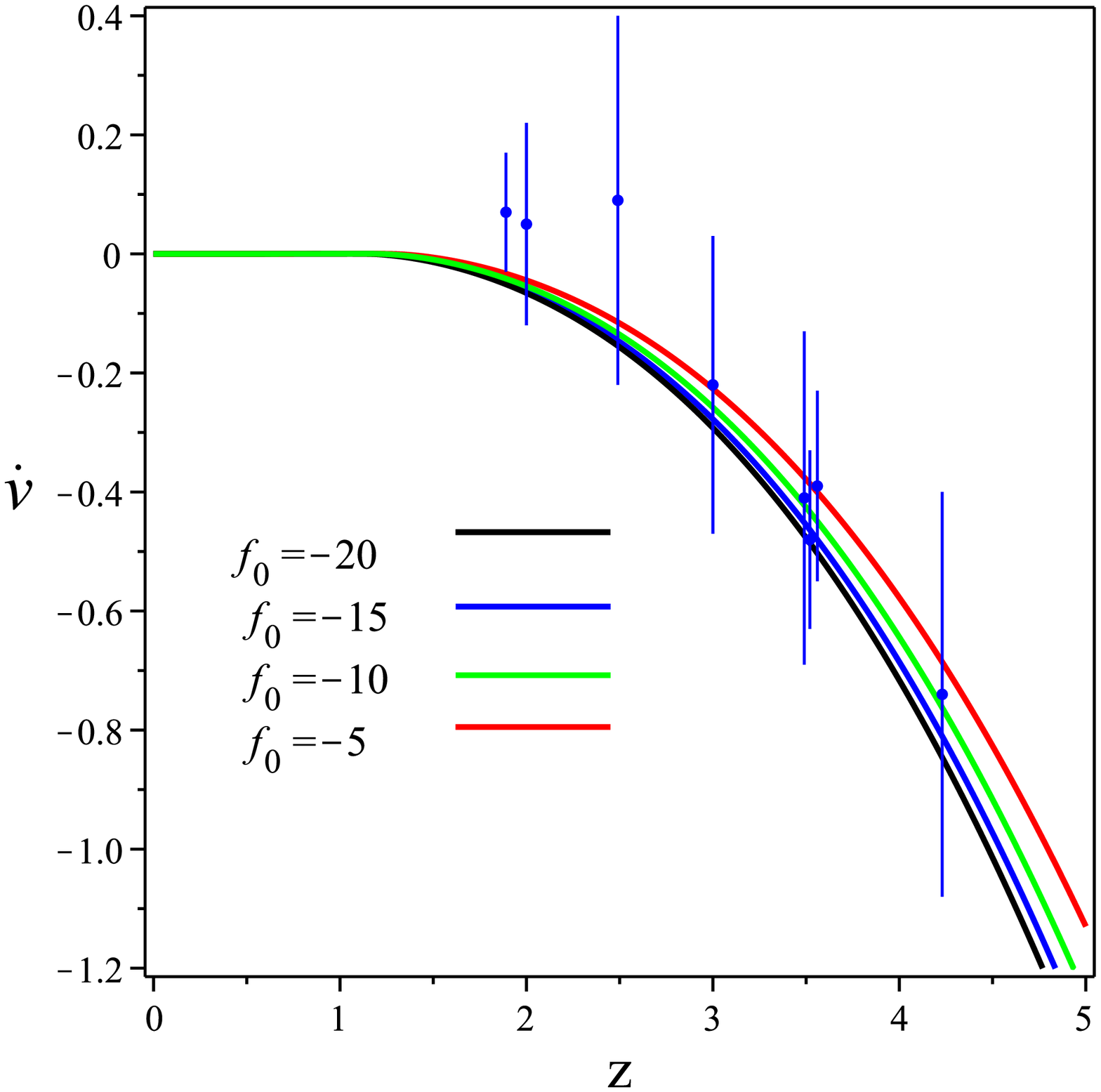}\hspace{0.1 cm}\\
Fig. 4: The graphs of the EoS parameter and velocity drift $\dot{v}$, plotted as functions\\
of $z$ for $\omega=\omega_{0}\phi(z)^{n}$, $n=-2$, $f=f_{0}e^{b\phi(z)}$, and $b = -1$
for four different value of $f_{0}$. \\
Initial values are $H(0)=1$, $\phi(0)=-10$, $\dot{\phi}(0)=3$.\\
\end{tabular*}\\

Alternatively, one may take a fixed $f_0$ and changes $\omega_0$ and numerically calculate the EoS parameter and velocity drift (see Fig.~5). \\

\begin{tabular*}{2.5 cm}{cc}
\includegraphics[scale=.4]{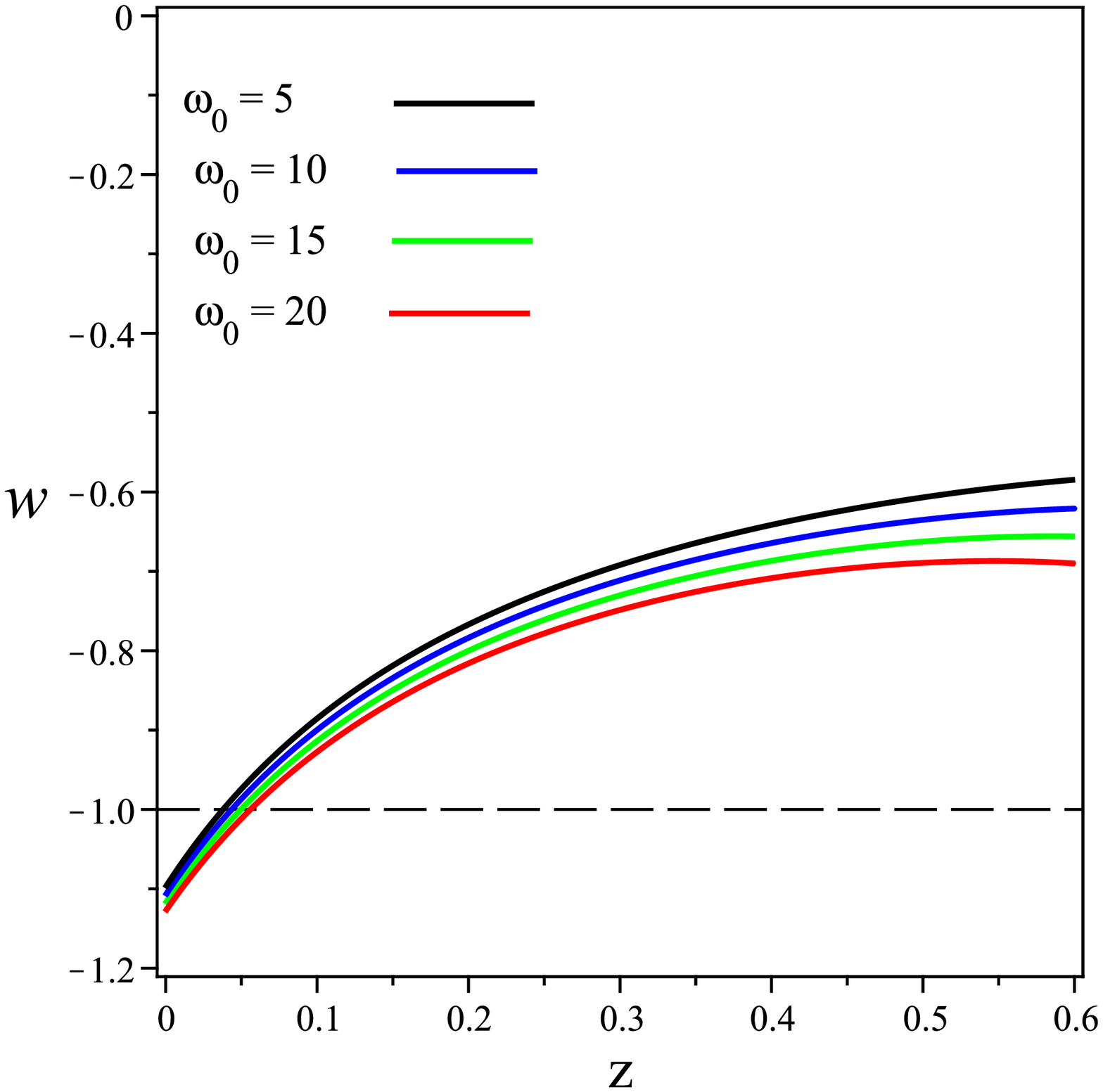}\hspace{0.1 cm}\includegraphics[scale=.4]{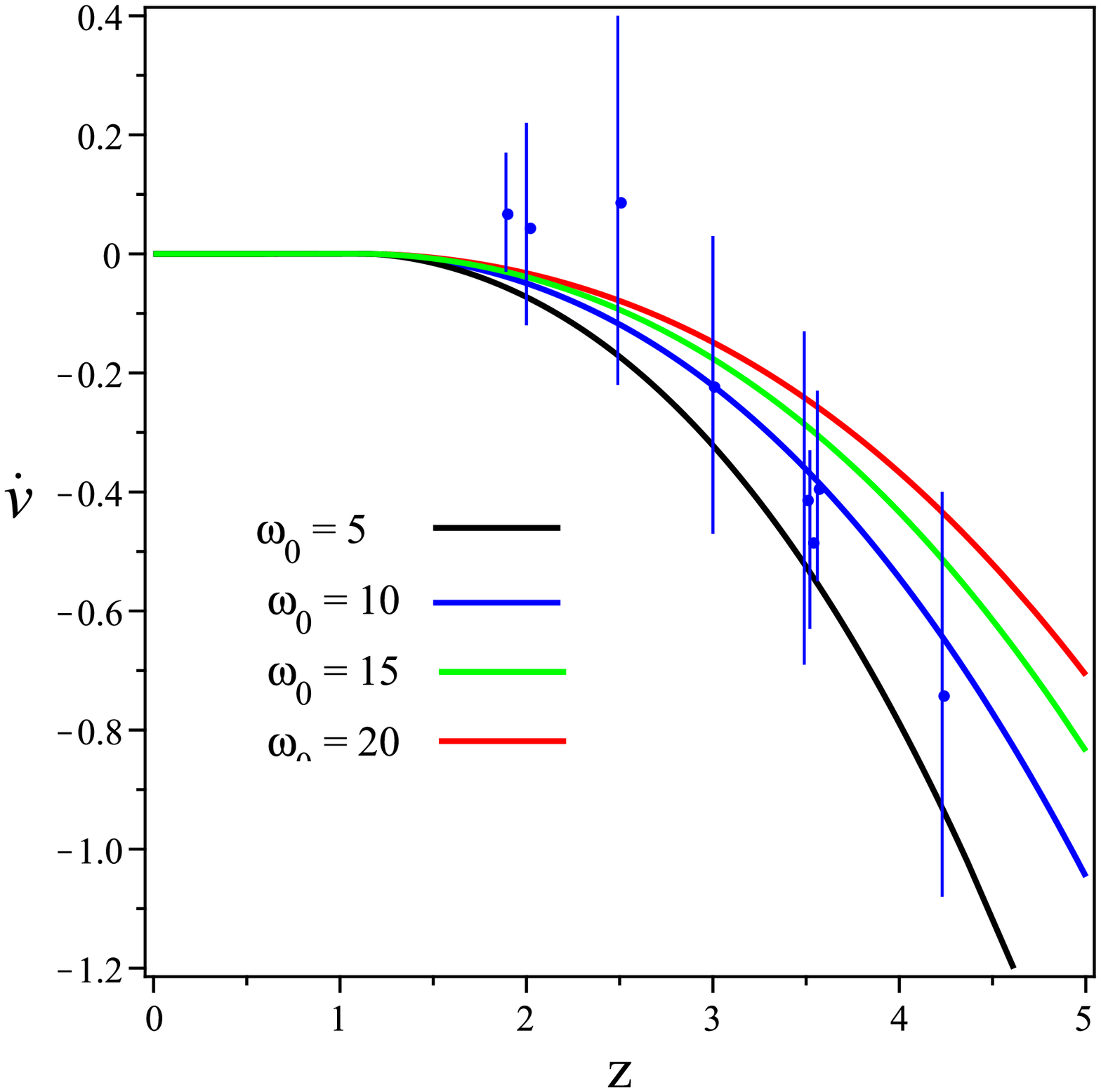}\hspace{0.1 cm}\\
Fig. 5: The graph of the scale factor $a$ and velocity drift $\dot{v}$, plotted as function\\
of $z$ for $\omega=\omega_{0}\phi(z)^{n}$, $n=-2$, $f=f_{0}e^{b\phi(z)}$, $f_0=-20$ and $b = -1$
for four different values of $\omega_{0}$. \\
Initial values are $H(0)=1$, $\phi(0)=-10$, $\dot{\phi}(0)=3$.\\
\end{tabular*}\\

In another attempt one may consider a case where both $f_0$ and $\omega_0$ change similar to $w_0$ and  $w_1$ in the CPL model.\\

\begin{tabular*}{2.5 cm}{cc}
\includegraphics[scale=.4]{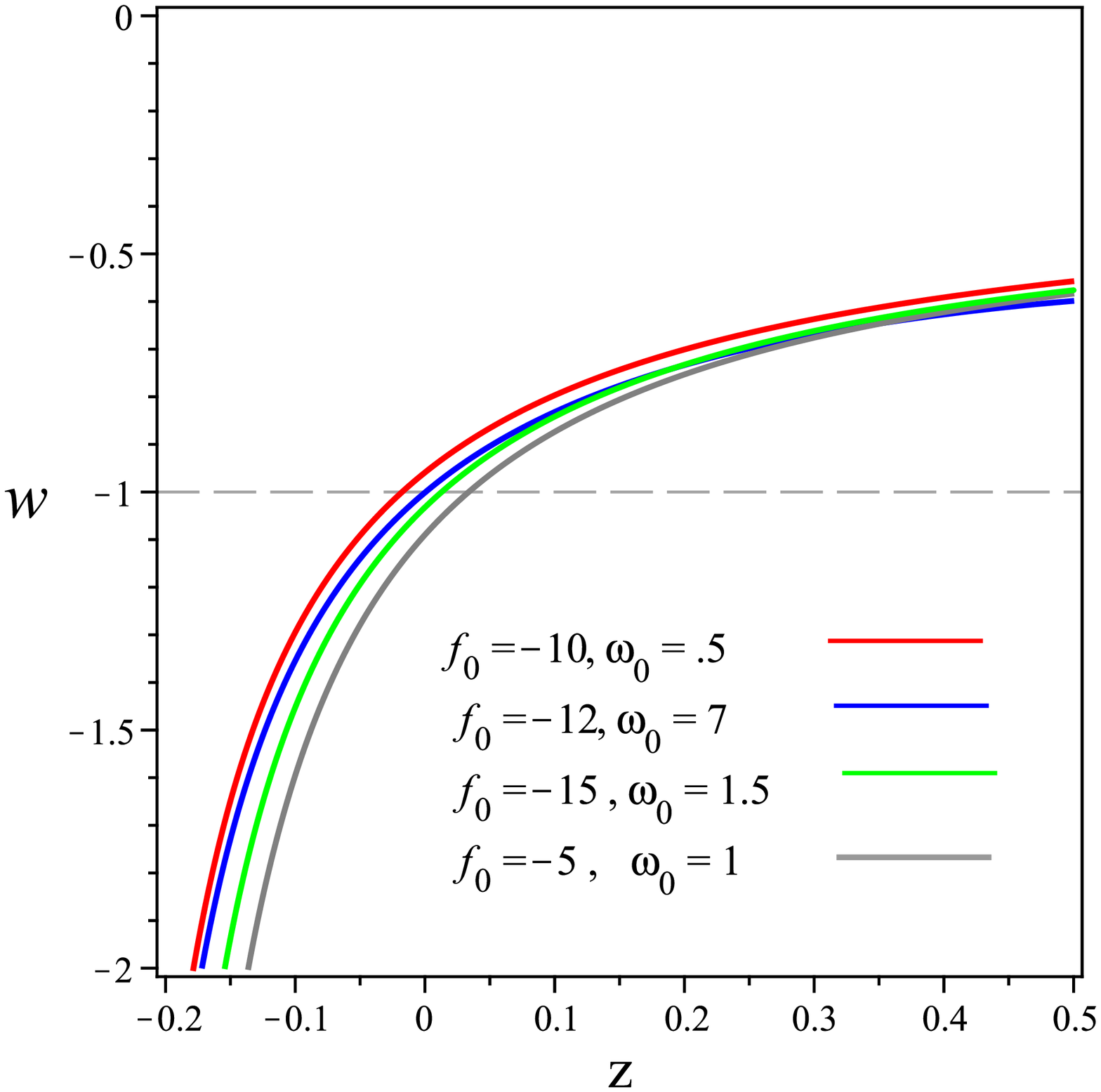}\hspace{0.1 cm}\includegraphics[scale=.4]{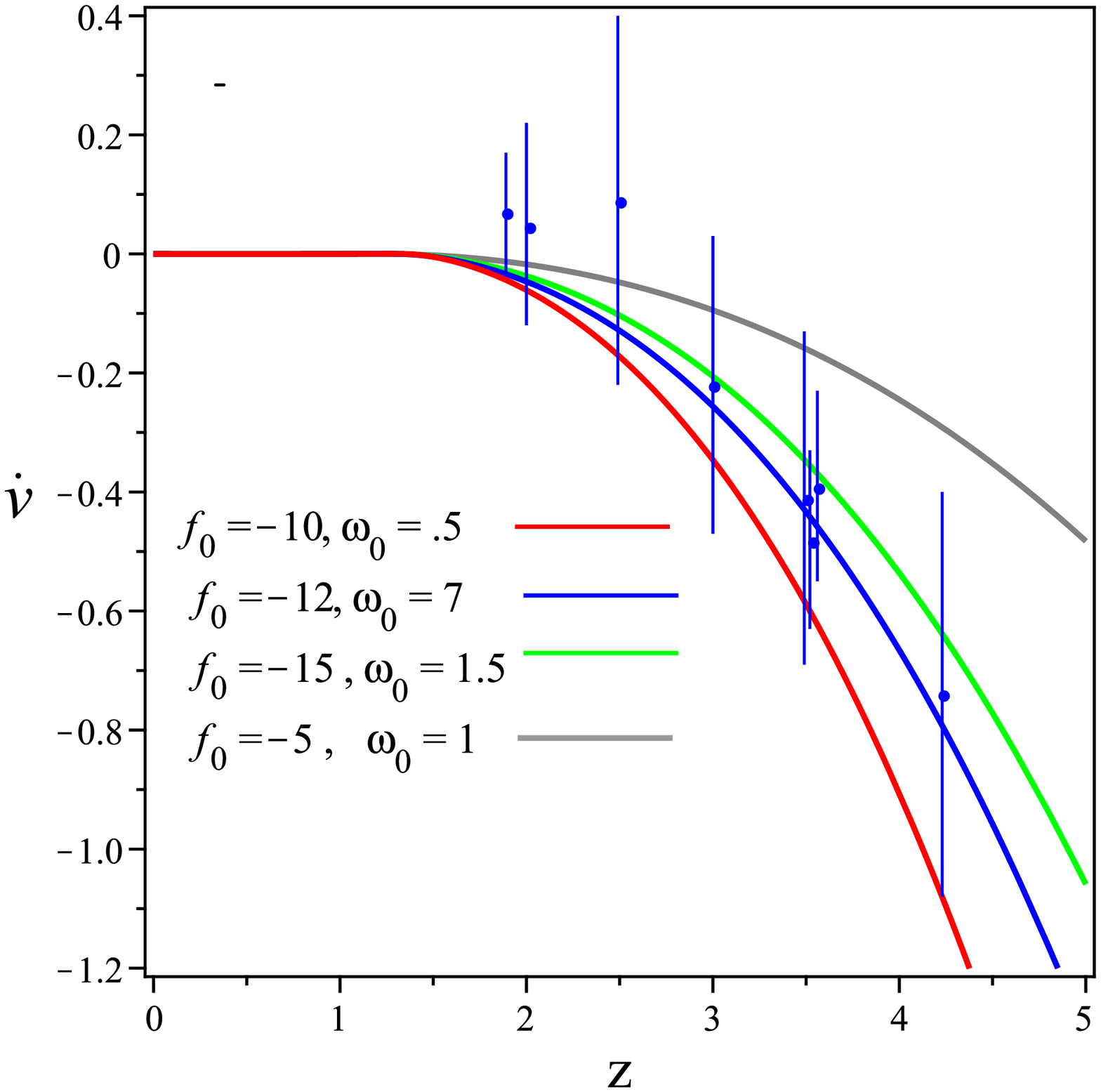}\hspace{0.1 cm}\\
Fig. 6: The graph of the scale factor $a$ and velocity drift $\dot{v}$, plotted as function\\
of $z$ for $\omega=\omega_{0}\phi(z)^{n}$, $n=-2$, $f=f_{0}e^{b\phi(z)}$ and $b = -1$
for four different values of $\omega_{0}$ and $f_{0}$. \\
Initial values are $H(0)=1$,  $\phi(0)=-10$, $\dot{\phi}(0)=3$.\\
\end{tabular*}\\

\subsection{The difference in the distance modulus, $\mu(z)$}

The difference between the absolute and
apparent luminosity of a distance object is given by, $\mu(z) = 25 + 5\log_{10}d_L(z)$ where the Luminosity distance quantity, $d_L(z)$ is given by
\begin{equation}\label{dl}
d_{L}(z)=(1+z)\int_0^z{\frac{dz'}{H(z')}}.
 \end{equation}
 In our model, from numerical computation one can obtain $H(z)$ which can be used to evaluate $\mu(z)$. To best fit the model for the parameters $f_{0}$ , $\omega_{0}$, $H_{0}$, $b$, $n$ and the initial conditions $\phi(0)$, and $\dot{\phi}(0)$ with the most recent observational data, the Type Ia supernovea (SNe Ia), we employe the $\chi^2$ method. We constrain the parameters including the initial conditions by minimizing the $\chi^2$ function given as
\begin{equation}\label{chi2}
    \chi^2_{SNe}(f_{0}, \omega_{0}, H_{0},b,n,\phi(0),\dot{\phi}(0))=\sum_{i=1}^{557}\frac{[\mu_i^{the}(z_i|f_{0}, \omega_{0}, H_{0},b,n,\phi(0),\dot{\phi}(0) - \mu_i^{obs}]^2}{\sigma_i^2},
\end{equation}
where the sum is over the SNe Ia sample. In relation (\ref{chi2}), $\mu_i^{the}$ and $\mu_i^{obs}$ are the distance modulus parameters obtained from our model and from observation, respectively, and $\sigma$ is the estimated error of the $\mu_i^{obs}$. In our model the best fit values occur at $f_{0}=-7$, $\omega_{0}=1.2$, $H_{0}=0.84$, $ b=-.4 $, $ n=-2 $, $ \phi(0)=1.5$, $ \dot{\phi}(0)=1$, with $\chi^2_{min}=804.87232$. In Fig. 7, the distance modulus, $\mu(z)$, in our model is compared with the observational data for the obtained parameters and initial conditions using $\chi^2$ method. Also in Fig. 8 the two dimensional likelihood distribution for model parameters $f_{0}$ , $H_{0}$ and $\omega_{0}$, $H_{0}$ are shown.\\

\begin{tabular*}{2.5 cm}{cc}
\includegraphics[scale=.44]{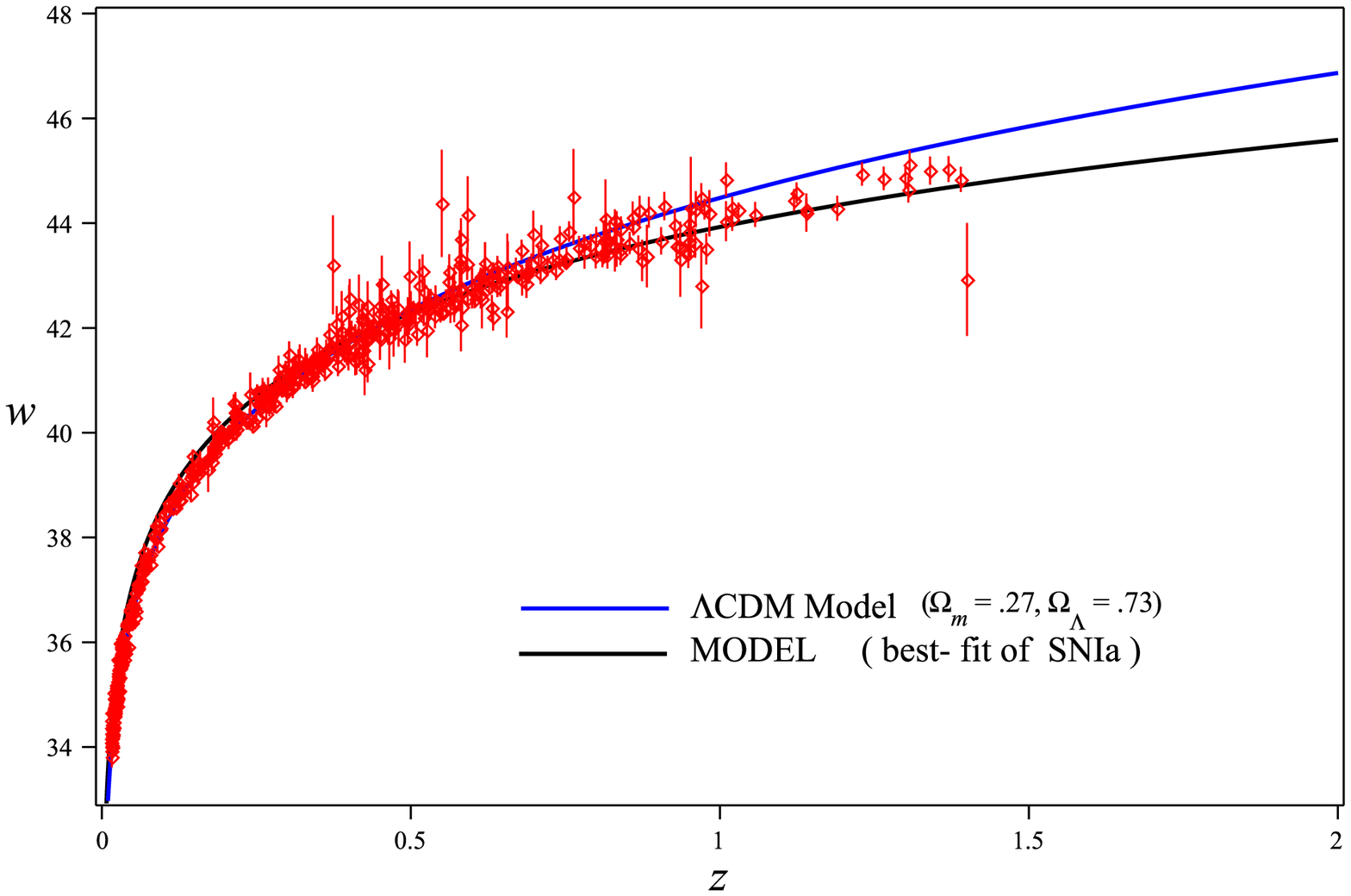}\hspace{0.1 cm}\includegraphics[scale=.44]{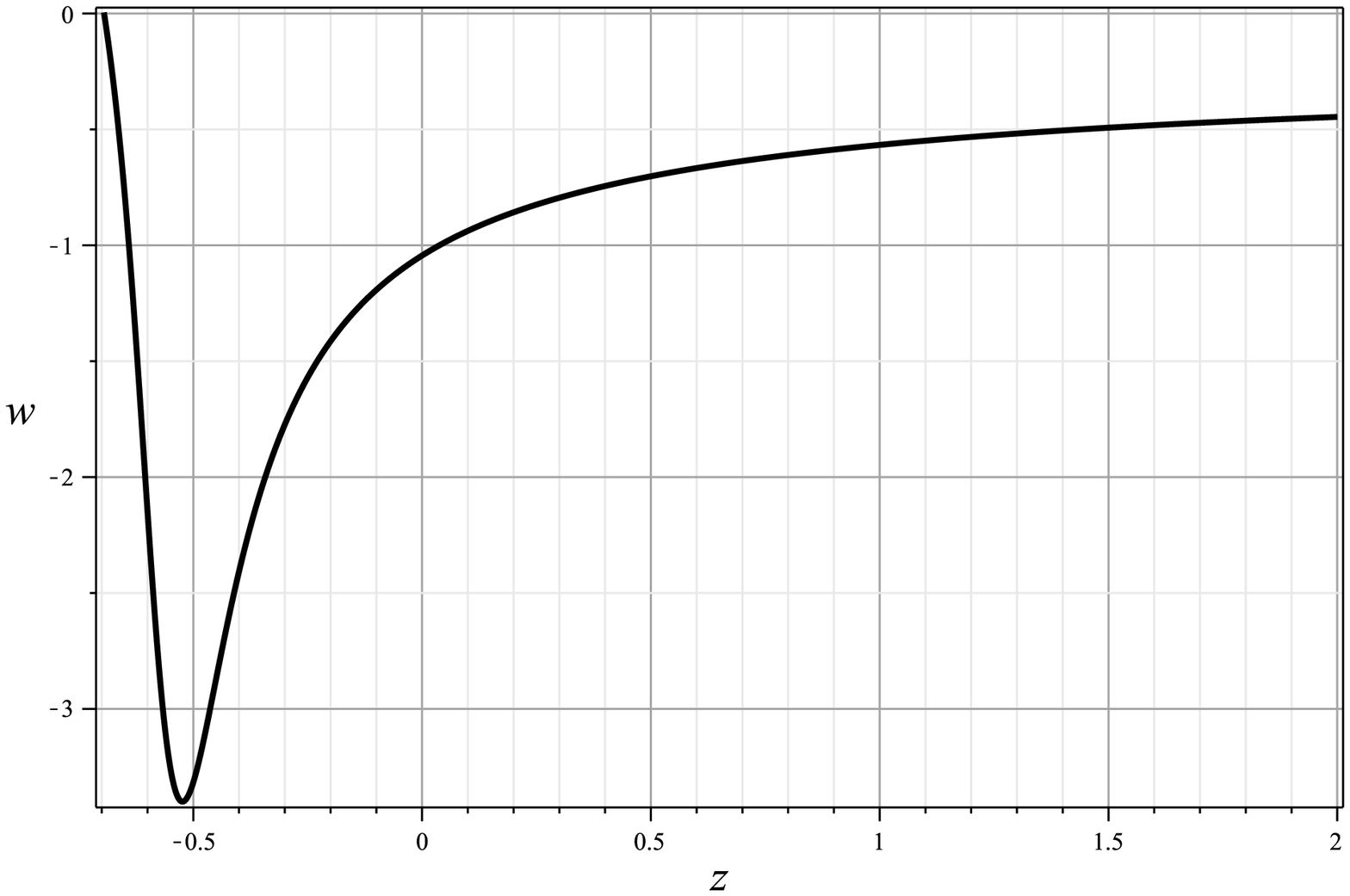}\hspace{0.1 cm}\\
Fig. 7: The graphs of the distance modulus $\mu(z)$ and EoS parameter $w$\\  plotted as function of the redshift for the best fitted model parameters
 \\$f_{0}=-7$, $\omega_{0}=1.2$, $H_{0}=0.84$, $ b=-.4 $, $ n=-2 $, and ICs. $\phi(0)=1.5$ $\dot{\phi}(0)=1$.\\
\end{tabular*}\\

\begin{tabular*}{2.5 cm}{cc}
\includegraphics[scale=.42]{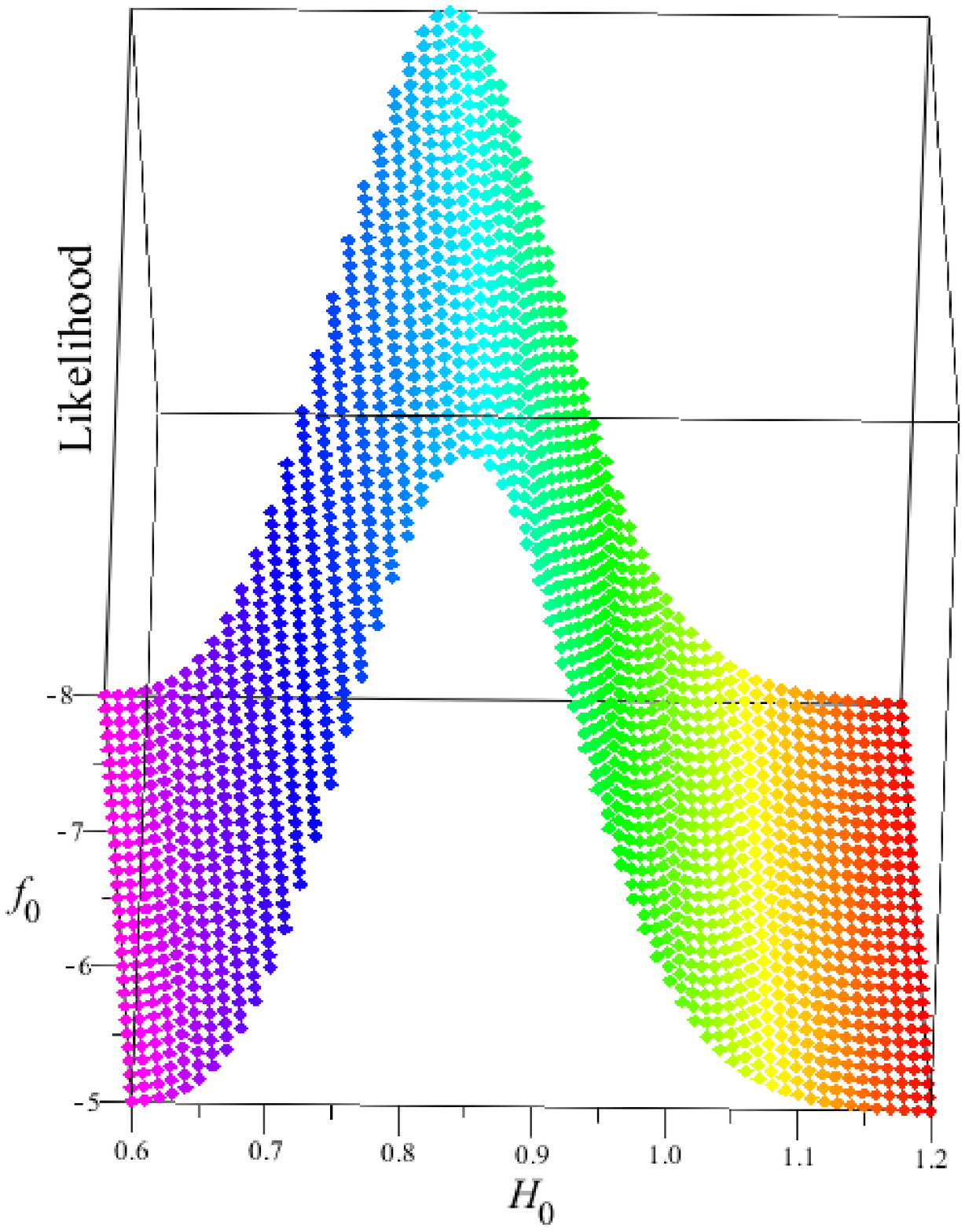}\hspace{0.1 cm}\includegraphics[scale=.4]{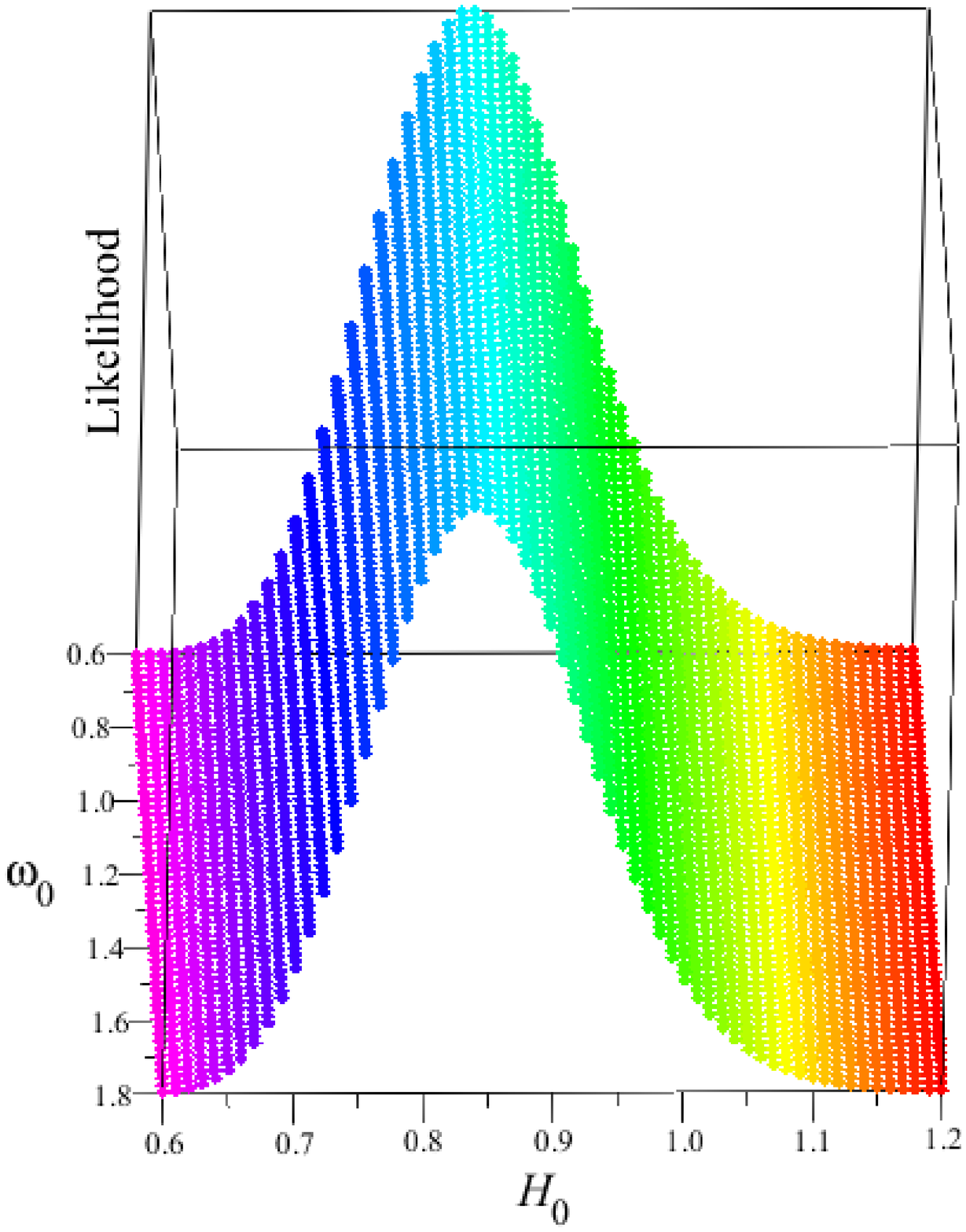}\hspace{0.1 cm}\\
 Fig. 8:  The graphs of two dimensional likelihood distribution for parameters $f_{0}$, $H_{0}$ and \\
$\omega_{0}$, $H_{0}$  with $b=-.4$, $n=-2$ and ICs. $\phi(0)=1.5$, $\dot{\phi}(0)=1$ and $d_L(0)=0$.\\
\end{tabular*}\\

\section{Results and Summary }

In this paper, we have studied the evolution of gravitational and scalar fields
in CGBD cosmological model in which a light scalar field
nonminimally coupled both to the matter Lagrangian (as chameleon field) and the scalar
curvature (as brans-Dicke field) in the action. The solution shows that the evolution
of the scale factor of the universe is non-singular in a bouncing
scenario, with an initial contracting phase which lasts until to
a non-vanishing minimal radius is reached and then smoothly
transits into an expanding phase. Also the dynamics of the EoS parameter with respect to time in early universe is
shown in Fig. 2 with a transition from phantom phase ( say phantom inflation) to non phantom phase towards matter dominated era.

We then analyze the model with the CRD test. The variation of
velocity drift for different values of $f_0$ and $\omega_0$ with
redshift $z$ is shown in Figs.~4, ~5 and ~6. In addition, the evolution of
the cosmological EoS parameter in terms of the redshift, with a transition from phantom phase ($w<-1$)
in the past to the non phantom phase ($w>-1$) in the recent past is favored
for values of $f_0$ and $\omega_0$, in agreement with the current
observational data. The PDL crossing occurs for different
values of $f_0$ and $\omega_0$ within the range of observationally
accepted redshift $z$.

A comparison between our
model with the CPL parametrization model shows that our model is in better agreement with
the data for values of redshift $z$ greater than $3$.
To best fit model parameters to data we need more redshift velocity drift data within the range
$z<2.5$. The PDL crossing in Figs.~4 and~5 shows that
in the former the crossing occurs for $-0.06<z<0.03$ whereas in the
latter for $0.03<z<0.06$. In regards to the drift velocity with
respect to the observational data it shows that both velocity drifts are in better agreement compare to the CPL model for
$z>3$ whereas the velocity drift against $z$ in Fig.~5 is even more satisfactory. Fig. 4 and 6 show that today and in near future the universe experience PDL crossing, in future it will be in  phantom phase whereas in near past was in non phantom phase. In comparison, Fig. 5 shows that in future and today it is in phantom phase, whereas PDL crossing occurred in very near past and in the past the universe was in non phantom phase.

In the second cosmological test, the difference in the distance modulus $\mu(z)$ is calculated for the model in comparison with the most recent observational data, the Type Ia supernovea (SNe Ia). To best fit our model with the data we use the $\chi^2$ and likelihood fitting test. The $\chi^2$ fitting in the model depends on the parameters $f_{0}$, $\omega_{0}$, $H_{0}$, $b$, $n$ and the initial conditions $\phi(0)$, and $\dot{\phi}(0)$. The multi-dimensional fitting analysis allows us to constrain the parameters that result in a minimum $\chi^2$. From Fig. 7 left), we observe that both our model best fitted distance modulus and $\Lambda CDM$ model are compared with the observational data. Although both models best fitted with the data for $z \leq 0.5$, our model better fitted for larger redshifts. In the Fig. 7 right) which the distance modulus is fitted with the most recent observational data by using $\chi^2$ method, we have a more observationally reliable EoS parameter. The graph shows that the current value of the EoS parameter is about $\sim 1.1$. In future the universe approaches phantom phase but later transit to the non phantom phase. In the near past the universe EoS parameter crossed the PDL, and approaches the non phantom phase in larger redshift. In Fig. 8, the two dimensional likelihood distribution for model parameters are given. The graphs show that, the model parameters are bounded in the regions $-8<f_{0}<05$, $0.6<\omega_{0}<1.8$ and $0.6<H_{0}<1.2$ with the best fit values are $-7$, $1.2$ and $.84$, respectively.

In figures 1-6 as can be seen, using different initial values for the model parameters, the trajectories eventually converge. So different initial values for the model parameters may initially cause divergence among the trajectories but eventually do not effect the dynamics. With respect to the initial conditions, comparing the graphs of EoS parameter in Figs. 4, 5 and 6 (with ICs. $\phi(0)=-10$, $\dot{\phi}(0)=3$), with the best fitted EoS parameter in graph 7 (with ICs. $\phi(0)=1.5$, $\dot{\phi}(0)=1$), one concludes that since the EoS parameters in different scenarios show similar behavior, the model is not sensitive to the initial conditions.

In summary, we may conclude that the CGBD model can predict the PDL crossing and fit the observational data for velocity drift and distance modulus better that the CPL and $\Lambda CDM$ models respectively, subject to constraints on the model parameters.

\end{document}